# Nanostructure and properties of a Cu-Cr composite processed by severe plastic deformation.

X. Sauvage[1]*, P. Jessner[1], F. Vurpillot[1], R. Pippan[2]

1- University of Rouen, Groupe de Physique des Matériaux, CNRS (UMR 6634), Avenue de l'Université - BP 12, 76801 Saint-Etienne du Rouvray, France
2- Erich Schmid Institute of Material Sciences, CD-Laboratory for Local Analysis of Deformation and Fracture, Austrian Academy of Sciences, Jahnstraße 12, A-8700 Leoben, Austria

**Abstract**

A Cu-Cr composite was processed by severe plastic deformation to investigate the role of interphase boundaries on the grain size reduction mechanisms. The as-deformed material exhibits a grain size of only 20nm. This gives rise to a dramatic increase of the hardness. Some deformation induced Cu super saturated solid solutions were clearly exhibited and it is shown that they decrease the hardness. The formation of such supersaturated solid solution and their influence on the mechanical properties are discussed.

*corresponding author: Xavier Sauvage
xavier.sauvage@univ-rouen.fr
Tel : + 33 2 32 95 51 42
Fax : + 33 2 32 95 50 32

More than fifty years ago, Hall and Petch have established the relationship between the grain size of metals and their macroscopic yield stress [1, 2]. Smaller grain size gives higher yield stress, thus nanostructuring metals and alloys looks very promising to reach record level of mechanical strength. Although the exact underlying mechanisms are still under debate, and especially for material with a grain size in the nanometre range [3], intensive research was developed during the past decade to bring nanostructured metals and alloys to applications. There are however two critical issues: the usual lack of ductility of nanostructured metals [4, 5], and the fabrication route to process bulk pieces. Numerous strategies have been developed to produce these materials such as electroplating, powder consolidation, or severe plastic deformation (SPD). Techniques related to this later approach seem extremely promising especially because materials are free of porosity. Pure metals or commercial alloys processed by SPD typically exhibit a grain size in a range 100 to 500 nm [6]. This feature is usually attributed to the minimum size of dislocation boundaries that can be achieved [7]. However, there are some exceptions and few SPD processed metallic alloys exhibit a grain size much smaller than 100 nm. This is the case of some intermetallics like $Ni_3Al$ and TiAl [8, 9], pearlitic steels [10, 11] or composites [12, 13]. To optimize grain refinement by SPD, a better understanding of the physical mechanisms involved in these specific cases is needed. The aim of this work was to clarify the role of interphase boundaries.

The material investigated in the present study is a Cu-Cr composite containing 43wt.% Cr and it was delivered by PLANSEE, Reutte (Austria). This system was chosen because these two elements can be considered as immiscible. The maximum solubility of Cu in bcc Cr is indeed less than 0.2 at.% and the maximum solubility of Cr in fcc Cu is 0.89 at% at 1350K, while there is no intermetallic phase reported in the phase diagram [14]. The initial microstructure of the composite investigated in the present study consists of bcc Cr particles with a mean diameter of about 50 μm embedded in a fcc Cu matrix (Fig. 1(a)). The volume fraction of Cr particle is about 50%. This material was processed by High Pressure Torsion (HPT) [6] at room temperature up to 25 revolutions under a pressure of 6 GPa and with a rotation speed of $2 \cdot 10^{-2}$ rad.s$^{-1}$. To quantify the evolution of the mechanical properties as a function of the applied strain, micro-hardness measurements were performed on a BUEHLER Micromet 2003 with a load of 100g. Microstructures were characterized by Scanning Electron Microscopy (SEM) using a LEO FE 1530 and Transmission Electron Microscopy (TEM) using a JEOL 2000 FX and a Philips CM12. TEM observations were performed at a distance of 3 mm from the HPT disc centre with the electron beam parallel to the torsion axis. Electron transparency of thin foils was achieved by ion milling using a PIPS-GATAN 691.

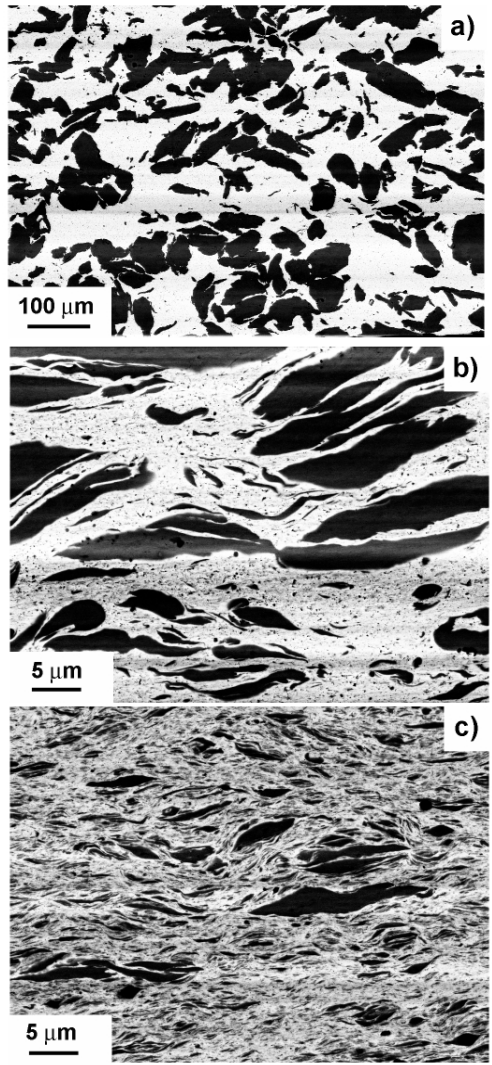

*Figure 1: SEM images of the microstructure of the CuCr43 composite in the cross section of HPT discs (The shear direction is horizontal; Cr particles are darkly imaged) : before deformation (a), deformed up to a shear strain $\gamma = 50$ (b), and up to a shear strain $\gamma = 150$ (c).*

Nano-analyses were carried out by Atom Probe Tomography (APT) using a CAMECA Tomographic Atom Probe detection system (TAP) [15]. Specimens were prepared using standard electropolishing methods [12] and field evaporated in UHV conditions with electric pulses (20% pulse fraction at 2kHz and specimen temperature of 50K) or femtosecond laser pulses (energy of $5 \, 10^{-7}$J at 2kHz and specimen temperature of 20K) [16]. Due to specimen failure during the field evaporation by electric pulses, volumes analysed this way are much smaller than those collected with the laser pulses. Thus electric pulses were used only as a reference to find out the optimal conditions in the pulsed laser mode.

The large shear strain applied to the original composite induces a significant elongation and necking of Cr particles in the cross section of HPT discs (Fig. 1 (b) and (c)). This mechanism induces an important reduction of their mean size down to few micrometers for a shear strain of $\gamma = 150$ (Fig. 1 (c)). This is one order of magnitude smaller than before deformation.

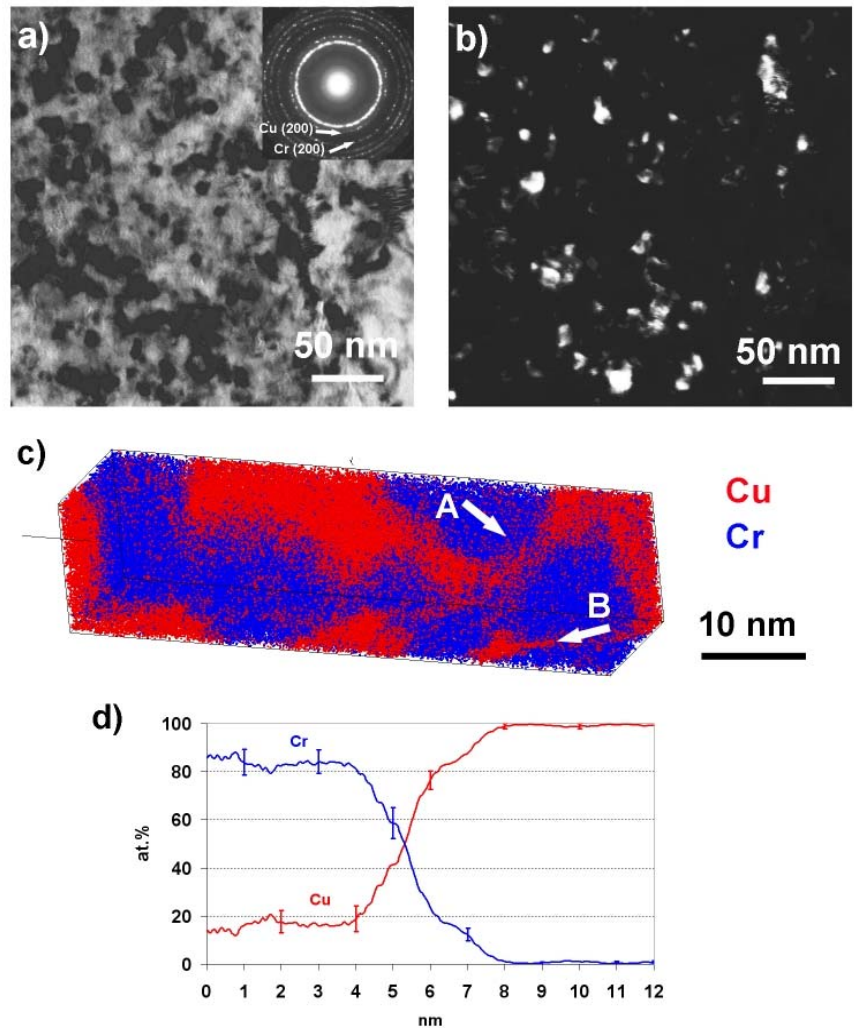

*Figure 2: (a) TEM bright field image and related SAED pattern of the CuCr43 composite processed by HPT up to 25 revolutions; (b) Dark field image of the same area obtained by selecting some (111)Cu and (110)Cr lattice reflections (first Debye-Scherrer ring); (c) 3D reconstruction of an analysed volume ($12 \times 12 \times 50$ nm$^3$) showing the distribution of Cu and Cr. The necking of a Cu grain (A) and a resulting nanoscaled tip (B) are arrowed (d) Composition profile computed across a Cu/Cr interface (sampling volume thickness 1nm).*

At higher level of deformation, the refinement of the microstructure continues. As shown on the TEM bright field image (Fig. 2(a)), after 25 revolutions by HPT (corresponding to a shear strain of about $\gamma = 700$ at 3 mm from the disc centre), the grain size is in the nanometre range. The selected area diffraction (SAED) pattern shows Debye-Scherrer rings characteristic of polycrystalline structures with a very small crystallite size. Rings corresponding to both to the fcc copper phase and to the bcc chromium phase are exhibited. The first ring results from the overlap of the Cr(110) and the Cu(111), but the two following rings correspond to Cu(200) and Cr(200) (marked with arrows on the SAED). The dark field image on the Fig. 2(b) was obtained by selecting a part of the first ring with a small aperture. Thus on this image both Cu and Cr grains are imaged. This image clearly shows that after 25 revolutions the grain size is in a range of only 10 to 20nm. This is one order of magnitude smaller than in commercially pure metals processed by HPT [6], indicating that interphase boundaries play a critical role in the grain size reduction mechanism. The atomic scale distribution of Cu and Cr atoms within this nanostructure was mapped out by APT. The Figure 2(c) clearly exhibits Cu and Cr domains with a size consistent with TEM observations. Some Cu grains are necked down to the nanoscale (marked A) and this feature leads to the formation of nanoscaled tip (marked B). On the image of this 3D volume, it seems that there is a significant amount of Cu in chromium grains. The composition profile computed across a typical Cu/Cr interface (Fig. 2(d)) confirms this point. In spite of the extremely low mutual solubility of Cu and Cr [14], the Cu concentration in bcc chromium grains is in a range of 10 to 20 at.% after SPD, while Cr atoms did not diffuse in the fcc copper phase.

Similar super saturated solid solutions (SSSS) in the Cu-Cr system have already been reported in ball milled powders [17, 18] or in sputtered thin films [19]. Shen and Koch have measured the heat release of as-milled $Cr_{80}Cu_{20}$ powders (10.5 kJ/mol), and of as-milled elemental Cr (4.2 kJ/mol) [18]. By subtraction, it gives an estimate of the enthalpy of formation of a 20%Cu SSSS in bcc Cr: 6.3 kJ/mol [18]. As shown on the Fig. 1 and Fig. 2, the shear deformation by HPT induces a strong necking of Cu grains. Thus following the model proposed by Yavari [20], small fragments with tip radii of the order of few nanometres appear (marked B on Fig. 2(c)). The free energy increases due to these tips writes as:

$$\Delta G = 2 V_m \sigma_{Cu/Cr} / r, \qquad (1)$$

where $V_m$ is the molar volume, $\sigma_{Cu/Cr}$ is the interfacial energy and r the tip radius. The bcc Cr and fcc Cu phases have a similar molar volume (about $7 \cdot 10^{-6}$ m$^3$), from Aguilar and co-authors the interfacial energy is 0.6 J m$^{-2}$ [17], thus from eq (1), $\Delta G$ overcomes the enthalpy of formation of the SSSS if the tip a radius is smaller than 1.3 nm.

Michaelsen and co-authors have shown that the enthalpy of formation of a Cr SSSS in fcc Cu is similar to that of a Cu SSSS in bcc Cr [19], however only this later case was detected by APT (Fig. 2(d)). This feature might be related to different diffusivity of Cu and Cr atoms within the nanostructure during SPD. The material is deformed under a high hydrostatic pressure and Cu atoms are smaller than Cr atoms. Under such conditions, the mobility of Cu atoms in bcc Cr would be certainly much higher than that of Cr atoms in the fcc Cu phase. This gives rise to the dissolution of Cu tips and the formation of a SSSS. Concerning Cr tips, they probably sphereodize by interface diffusion to minimize capillary pressures. These two mechanisms associated with necking give rise to the nanoscaled equi-axed structure exhibited after 25 revolutions of HPT (Fig. 2). However one should note that the material was deformed at room temperature where the atomic mobility of both Cu and Cr is extremely limited [19]. The SSSS analysed by APT in the present study are homogeneous and interfaces with fcc Cu grains are quite sharp (Fig. 2(d)). This indicates that even if some dislocations could act as diffusion pipe, the atomic mobility might be mostly enhanced by SPD induced vacancies [21].

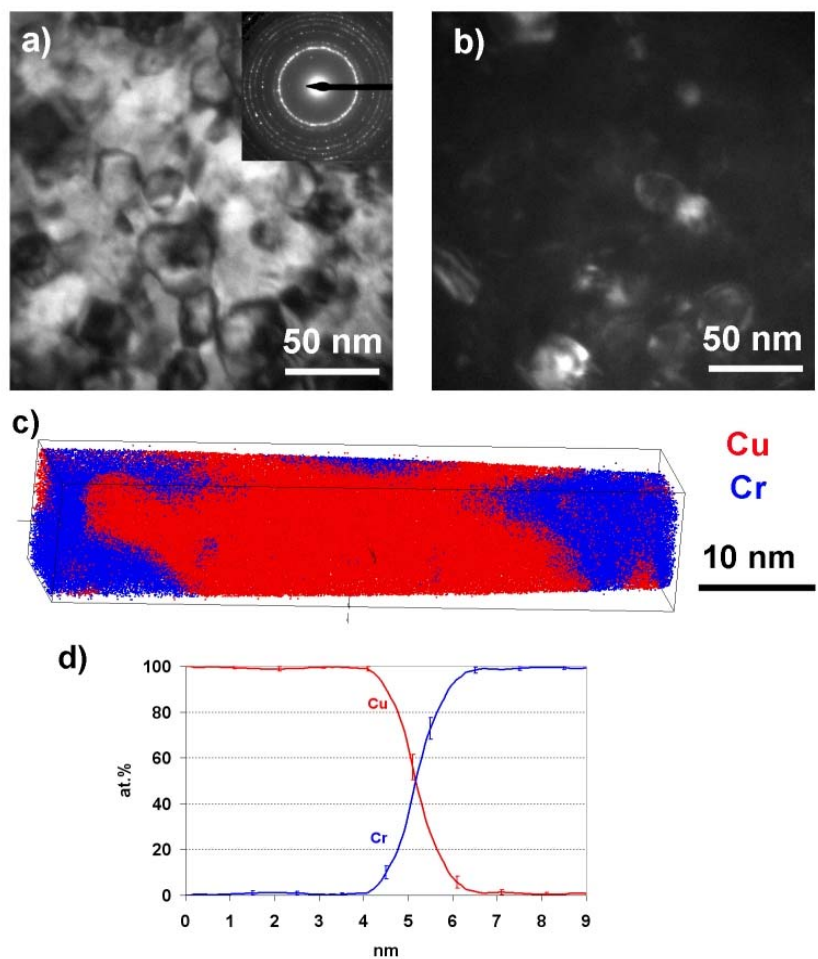

*Figure 3: (a) TEM bright field image and related SAED pattern of the CuCr43 composite processed by HPT up to 25 revolutions and aged during 30min at 450°C; (b) Dark field image of the same area obtained by selecting some (111)Cu and (110)Cr lattice reflections (first Debye-Scherrer ring); (c) 3D reconstruction of an analysed volume (10x10x50 nm$^3$) showing the distribution of Cu and Cr ; (d) Composition profile computed across a Cu/Cr interface (sampling volume thickness 1nm).*

Previous studies on ball milled powders or sputtered thin films have demonstrated that phase separation in Cu-Cr SSSS starts in a temperature range of 400 to 500°C. Thus, the present material processed by HPT up to 25 revolutions was aged during 30min at 450°C. TEM observations clearly show that some significant grain growth occurred during ageing and the grain size is now in a range of 20 to 40 nm (Fig. 3 (a) and 3 (b)). This grain growth is also clearly exhibited on the image of the 3D volume analysed by APT (Fig. 3(c)), and the composition profile computed across a Cu/Cr interface demonstrates also that chromium grains are free of Cu after aging. Thus, phase separation occurred during aging at 450°C as expected.

*Table 1: Microhardness, grain size range and Cu concentration of bcc Cr of the CuCr43 composite measured at a distance of 3mm from the disc centre as a function of the number of revolution (N). Data of the sample processed by 25 revolutions and then aged during 30 min at 450°C are also reported for comparison.*

| N | 0 | 5 | 16 | 25 | 25 (+ 30 min 450°C) |
|---|---|---|---|---|---|
| **HV (GPa)** | 0.8 ± 0.1 | 2.7 ± 0.1 | 4.3 ± 0.1 | 3.8 ± 0.1 | 4.3 ± 0.1 |
| **Grain size range** | 40 – 60 μm | 0.5 – 3 μm | 10 – 20 nm | 10 – 20 nm | 20 – 40 nm |
| **% of Cu in bcc Cr** | --- | --- | 7 ± 2 at.% | 15 ± 5 at.% | < 1 at.% |

The evolution of the hardness and of the grain of the Cu-Cr composite as a function of the number of revolution by HPT is reported in the table 1. The SPD induced grain size reduction clearly leads to a strong hardening (from 0.8 GPa before HPT, up to 4.3 GPa after 16 revolutions). However, while there is a saturation of the grain size for a number of revolutions larger than 16, the hardness decreases from 4.3 down to 3.8 GPa. One should note that the grain size saturates but the amount of Cu in SSSS in bcc Cr after 25 revolutions (about 15 at.%) is twice as high as after 16 (about 7 at.%). This suggests that the bcc Cr phase is softened by Cu SSSS. The hardness of nanocrystalline Cu and Cr with a grain size in a range of 10 to 20 nm has been reported by Shen and Koch [18]. It is 13 GPa and 2 GPa for Cr and Cu respectively. The hardness of the nanostructured composite could be estimated by a simple rule of mixture. A Cr volume fraction of 50% should give a hardness of about 7 GPa, which is twice as large as the measured value. This discrepancy could be attributed to the formation of SSSS that were detected only for the highest level of deformation (16 or 25 revolutions). The hardness measured after ageing support also this idea. Indeed, although the grain size is twice larger than in the as-deformed material, the hardness is 0.5 GPa higher (Table 1). One should note that a similar solid solution softening effect has been reported by Shen and Koch in ball milled powders [18]. Thus, these experimental data indicate that the deformation mechanisms in such multiphase nanoscaled systems could be significantly affected by non-equilibrium composition gradients.

In conclusion, TEM observations and APT analyses clearly demonstrate that a bulk Cu-Cr composite with a grain size in a range of only 10 to 20 nm could be achieved by HPT processing. It demonstrates that interphase boundaries (for instance Cu-Cr boundaries) play a critical role in the grain size reduction mechanism. During SPD, some Cu diffuses in bcc Cr grains giving rise to SSSS. Capillary pressures are though to be the he driving force of this reaction. Such SPD induced SSSS significantly soften the nanostructured Cu-Cr composite.

**Acknowledgement**

*The authors gratefully acknowledge PLANSEE AG for supplying the material investigated in the present study.*